\documentstyle[epsfig]{mn}

\def\be{\begin{equation}}
\def\ee{\end{equation}}

\def\etal{{\it et al.~}}
\def\gtsima{$\; \buildrel > \over \sim \;$}
\def\ltsima{$\; \buildrel < \over \sim \;$}
\def\prosima{$\; \buildrel \propto \over \sim \;$}
\def\gsim{\lower.5ex\hbox{\gtsima}}
\def\lsim{\lower.5ex\hbox{\ltsima}}
\def\simgt{\lower.5ex\hbox{\gtsima}}
\def\simlt{\lower.5ex\hbox{\ltsima}}
\def\simpr{\lower.5ex\hbox{\prosima}}

\def\beq#1{\begin{equation}\label{#1}}
\def\eeq{\end{equation}}
\def\beqa#1{\begin{eqnarray}\label{#1}}
\def\eeqa{\end{eqnarray}}

\def\ie{{\frenchspacing\it i.e. }}
\def\eg{{\frenchspacing\it e.g. }}

\title[Cosmological Origin of the Lowest Metallicity Halo Stars]
{Cosmological Origin of the Lowest Metallicity Halo Stars}
\author[X. Hernandez and A. Ferrara]
{Xavier Hernandez$^1$ and Andrea Ferrara$^{1,2}$\\
$^1$ Osservatorio Astrofisico di Arcetri, Largo E. Fermi 5, 50125 Firenze, Italy\\
$^2$ Center for Computational Physics, University of Tsukuba, Tsukuba-shi,
Ibaraki-ken, 305-8577, Japan\\
}

\date{\today}

\begin{document}
\maketitle

\begin{abstract}

We explore the predictions of the standard hierarchical clustering
scenario of galaxy formation, regarding the numbers and metallicities of
PopIII stars likely to be found within our Galaxy today. By PopIII we shall be referring
to stars formed at large redshift ($z>4$), with low metallicities ($[Z/Z_{\odot}]<-2.5$) 
and in small systems (total mass $\simlt$ $2\times 10^{8} M_{\odot}$) that are 
extremely sensitive to stellar feedback,  and which through
a prescribed merging history (Lacey \& Cole 1993) end up becoming   part of the Milky
Way today. An analytic, extended Press-Schechter formalism is used to get the mass functions
of halos which will host PopIII stars at a given redshift, and which
will end up in Milky Way sized systems today. Each of these is modeled as a mini
galaxy, with a detailed treatment of the dark halo structure, angular momentum distribution,
final gas temperature and disk instabilities, all of which determine the fraction 
of the baryons which are subject to star formation. Use of new primordial metallicity 
stellar evolutionary models allows us to trace the history of the stars formed, give
accurate estimates of their expected numbers today, and their location in $L/L_{\odot}$ vs.
$T/K$ HR diagrams. A first comparison with observational data suggests that the IMF of
the first stars was increasingly high mass weighted towards high redshifts, levelling off
at $z\simgt  9$ at a characteristic stellar mass scale $m_s=10-15 M_\odot$. 

\end{abstract}


\section{INTRODUCTION}

The formation and the properties of the first stars in the universe are probably among
the most fascinating problems in present-day cosmology. Numerical simulations and analytical 
studies in a cosmological context (Tegmark \etal 1997, Abel \etal 1998, 
Bromm, Coppi \& Larson 2000, 
Machacek, Bryan \& Abel 2000) have reached a consensus on the fact that
the fragments potentially leading to star formation were already available inside 
very small dark matter halos at redshift $z\approx 30$ in most cosmological models. After that
point the requirements concerning the dynamic range and number of physical processes 
to be treated become so demanding that
one has to turn to simplified and/or lower dimensional studies.  
Although considerably refined, these studies (Omukai \& Nishi 1999,
Susa \& Umemura 2000, Nakamura \& Umemura 2000,
Ripamonti \etal 2000) have not yet reached an agreement on the mass of the 
stellar seed (core) and on
the accretion rate by which the former grows into  a star. Determining if and when the accretion   
is quenched seems to hold the key for the understanding of the properties of the first stars. 
However, the large variety  of physical effects to be considered (for a review see Ferrara 2000) 
again complicate the problem tremendously.  

On general grounds some authors have speculated that the IMF of the first stars can 
be skewed towards high stellar masses (Larson 1998) or even be bimodal (Nakamura \& Umemura 2000). 
One of the consequences of this assumption is an increase of the relative number of
massive, short-lived stars. These stars    will soon end their life as supernovae, leaving
either a compact remnant (a neutron stars or a black hole) or nothing at all (for masses
above $ \approx 120 M_\odot$, see Umeda \etal 2000 and Fryer, Woosley \& Heger 2000),
 and possibly releasing detectable 
bursts of gravitational waves (Schneider \etal 2000). In turn, this implies a strong decrease of the 
amount of zero- or very low-metallicity stars presumably  included in the
present-day Milky Way (MW), according to the hierarchical paradigm.  
Given the above discussion it appears that number counts of very low-metallicity 
 MW stars can not only 
provide evidences for ``living fossils", \ie stars formed immediately
 after the end of Dark Ages, but also
probe the primordial IMF and hierarchical models to cosmic epochs ($z\simgt 6$) currently 
unreachable to essentially all experiments.

We propose here that the first stars are born inside the so-called PopIII objects 
(Haiman \& Loeb 1996, Gnedin \& Ostriker 1997, Ciardi \etal 2000)
\ie systems with total mass $\simlt  2\times 10^8 M_\odot$, which strongly rely 
on molecular hydrogen cooling in order to collapse, given their low virial temperatures; 
in short, we define as PopIII stars those formed inside such mini galaxies. As Ciardi \etal
have pointed out, these objects are extremely fragile to the 
energy release by their own supernovae;
as a result, their gas is quickly evacuated and they can only witness a single
burst of star formation.  The remnant is a "naked stellar object" (Ciardi \etal 2000), 
\ie a tiny agglomerate  of long lived low-mass stars, essentially retaining the  
metallicity of the gas from which they formed. These are the stars that 
we are considering as candidates for the very low-metallicity stars observed in the MW halo. 

Our paper is structured as follows: the next section presents the 
Press-Schechter formalism used to calculate the mass functions
of MW sub-galactic clumps at different redshifts. In Section 3 we describe the treatment
of PopIII halos  as mini galaxies to estimate the number of stars each contributes to
the final MW. Section 4 gives a comparison with recent observational results  
of the numbers and metallicity distribution of extremely low metallicity halo stars. 
Finally, a brief summary is given in Section 5. 

\section{BUILD UP OF THE MILKY WAY}

We use the extended Press-Schechter formalism to calculate the conditional probability
that a virialized halo of mass $M_{1}$ at $z_{1}$ will become part of a more
massive halo of mass $M_{0}$ at a later time, $z_{0}$. This 
allows us to calculate the mass functions
of the fragments which through mergers will end up as part of a MW sized system today.
Following Nusser \& Sheth (1999), we identify the mass contained in progenitors of mass
between $M_{1}$ and $(M_{1}+dM_{1})$ with:
\be
\Delta M={M_{0}\over{\sqrt{2 \pi}}} {\delta_{c}(z_{1}-z_{0})\over{(S_{1}-S_{0})^{3/2}}}
exp\left[-{ {\delta_{c}^{2}(z_{1}-z_{0})^{2}}\over{2(S_{1}-S_{0})} } \right]
\left| {dS_{1}} \over {dM_{1}} \right| dM_{1}
\ee
\begin{figure}
\psfig{file=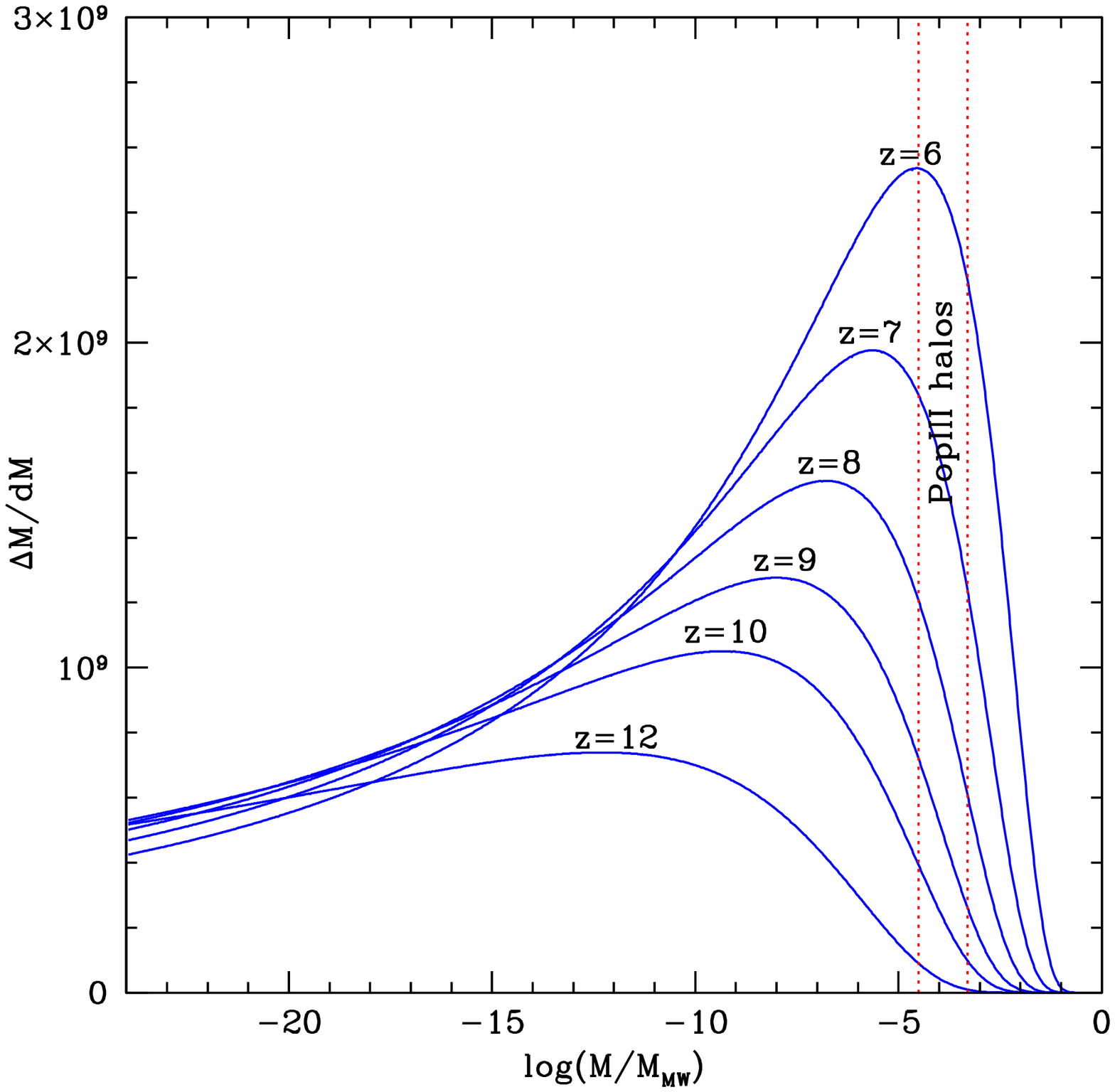,angle=0,width=9.0cm}
\caption{Mass functions of halos which at $z=0$ end up as part
 of a $2\times 10^{12} M_{\odot}$ system, i.e. similar to the MW,
at different redshifts. The dotted lines indicate the mass range where PopIII stars
are expected to form. }
\label{fig1}
\end{figure}
In the above expression $\sqrt{S_{i}}$ is the rms density fluctuation in a top hat window
function of radius $(3M_{i}/4 \pi \rho_{0})^{1/3}$, and $\delta_{c}$ the critical overdensity for
collapse, with $\rho_{0}=3\Omega_{M} H^{2}_{o}/8 \pi G$ being
the present mean mass density of the universe. 
Both of the above quantities were calculated using a $\Lambda CDM$ power
spectrum normalized to $\sigma_{8}=0.91$, in a $\Omega_{\Lambda}=0.7$ flat universe 
($\Omega_{M}=0.3$),
with $H_{0}=65$ km s$^{-1}$Mpc$^{-1}$, which we used throughout, 
\eg Percival \& Miller(1999). A baryonic fraction of 
0.05 of the total halo mass was assumed for all halos.

Current direct estimates of the mass of the MW rely on distant galactic satellites
used as probes of the mass internal to their current positions, using their
line of sight velocities and reasonable assumptions on their orbits. These methods
give values for the total mass of the MW of around $1 \times 10^{12} M_{\odot}$,
\eg  Kulessa \&  Lynden-Bell (1992), Binney \& Merrifield (1998). 
However, those estimates yield only lower limits,
as the total extent of the Galactic dark halo could be greater than the current positions
of suitably observable satellites. Consistently with the Press-Schechter approach we are 
using, we estimate the total mass of the MW        as the total mass contained within a
sphere of radius $r_{200}$, the radius inside which the average density is 200 times
the present background density, $\rho_{200}= 200 \times \rho_{0}$.
This in fact is the only meaningful definition of ``galaxy'' within the Press-Schechter approach. 
The circular velocity
at this radius would be around $V_{max}=\sqrt{2} \times V_{200}$, 
where $V_{max}=220$~km~s$^{-1}$ and the factor of $\sqrt{2}$ between $V_{max}$ 
and $V_{200}$ is deduced 
from CDM simulations \eg Avila-Reese \etal (1999), Steinmetz \& Navarro (1999). 
This gives:
\be
M_{MW}=\left( 3 \over 4 \pi \right)^{1/2} {V_{200}^{3} \over {G^{3/2} \rho_{200}^{1/2}}}
\ee
which for the adopted cosmological model gives 
$2.6 \times 10^{12} M_{\odot}$, somewhat above the observational lower limits. In what follows
we shall use $M_{MW}=2.0 \times 10^{12} M_{\odot}$, within the uncertainties associated with
going from  $V_{max}$ to $V_{200}$, and in determining $V_{max}$. 

\begin{figure}
\psfig{file=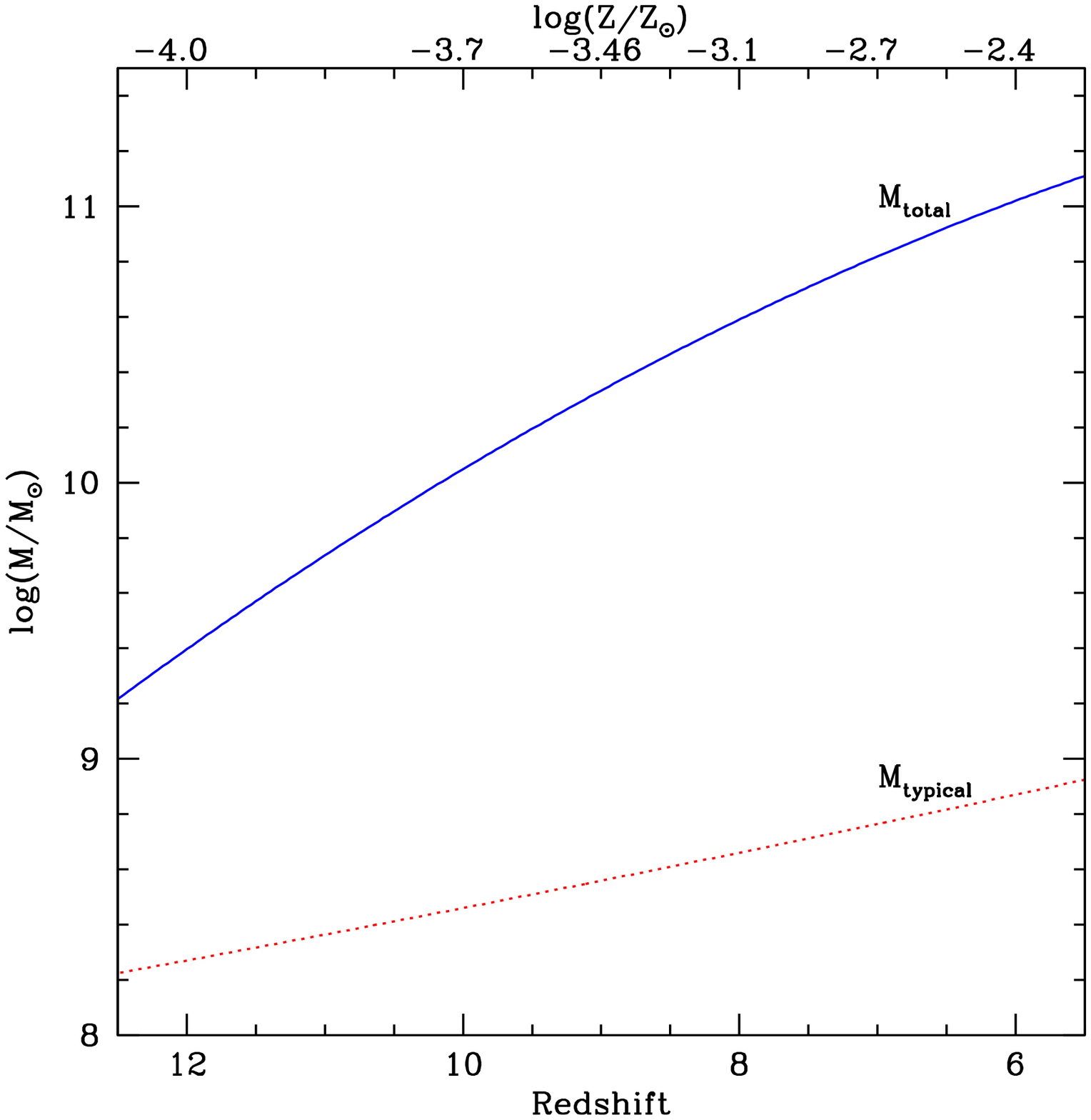,angle=0,width=9.0cm}
\caption{Total mass in PopIII halos that ends up in a MW       
sized galaxy  today, as a function
of redshift (solid line). The dotted curve shows the typical mass of these PopIII halos. 
The upper axis shows the average metallicity of the IGM at that redshift.
}
\label{fig2}
\end{figure}

Fig. 1 shows the mass of sub-halos which at $z=0$ find themselves included in
MW sized halos, as a function of sub-halo mass, at different redshifts; 
the dotted lines show the limits within which halos at the redshifts 
shown will form PopIII halos. 

We associate a maximum metallicity to the
stars to be formed and contained within 
these halos through the mean mass weighted metallicity
of the universe at each redshift, as calculated by Gnedin \& Ostriker (1997) using fully
hydrodynamic simulations. As a result of the merger process, these halos will probably contain
a few stars with yet lower metallicities formed in halos of the preceding hierarchies; in
this sense we take the metallicities of Gnedin \& Ostriker (1997) as an upper limit for the
stars of these halos.

The lower mass limit, $M_{crit}$, is taken from Tegmark \etal (1997) (see however Fuller \& 
Couchman 2000). Systems having virial temperatures below this curve cannot cool in a Hubble time. 
The upper limit, $M_{by}$, corresponds to the upper mass at which  
the stellar feedback due to supernova can expel the baryonic component entirely, thus
quenching further star formation. Following Ciardi \etal (2000), we define this as a
"blowaway", and we adopt their  determination for the evolution of $M_{by}$. Thus, 
larger systems where self enrichment can continue and stars of
metallicities higher than the Gnedin \& Ostriker (1997) limit will be formed are excluded. 
However, given
the sharply decreasing mass functions, our results are rather insensitive to this upper
limit. These two limits change slightly with redshift, the ones shown are typical for the
redshifts displayed, although the more detailed redshift dependent estimates discussed above 
were used.

Fig. 2 shows the integral of the PopIII halos mass functions between the selected limits,
$M_{total}$, as a function of redshift. The upper axis
shows the mean mass weighted metallicity corresponding to each redshift, discussed above. 
Fig. 2 also gives the typical mass of a single PopIII
halo, $M_{typical}$, showing that within the redshift interval displayed, each present day 
MW sized system will have had between $10$ and $100$ PopIII halos as progenitors, contributing
stars with metallicities in  $-4.0 < [Z/Z_{\odot}] < -2.4$. 
This means that the mass functions of
eq. 1 will be sampled densely, and the spread between different present day systems will 
be minimal i.e., all MW sized systems should contain the same numbers and distribution of stars
having metallicities in the above range. The two curves of Fig. 2 intersect at a 
redshift of $\sim 15$, corresponding to a metallicity of $[Z/Z_{\odot}]=-4.3$ in the 
Gnedin \& Ostriker picture. At this limit many present day MW sized systems will have had only
one PopIII halo coming from beyond this redshift, and many will have had none at all. In this
way we can expect that the numbers of stars in the Milky Way with metallicities below 
$[Z/Z_{\odot}]=-4.3$ should be minimal, indeed, no such stars have been found to date.

For even lower metallicities, one goes to
larger limiting redshifts to reach $z \sim 30$, $Z/Z_{\odot}\approx 0$ and 
$M_{typical} \sim 10^{6} M_{\odot}$, the objects responsible for the appearance of the very 
first stars and of starting the reionization of the universe. However, as commented above,
the chance of finding any such object as part of the MW is slim, they will be found 
as part of larger systems, perhaps as intracluster stars. For simplicity though, we will
use the term first stars to identify the earliest stars of our Galaxy.

\section{FORMING THE FIRST STARS}

\subsection{Disk Formation}

We model each of the PopIII halos discussed above as a small galactic system. Systems
with masses larger than $M_{crit}$ cool and the
gas forms a disk in centrifugal equilibrium (often   seen in the numerical 
simulations quoted above), with an exponential
surface density profile fixed by the total baryonic mass and angular momentum. This last
is given by the $\lambda$ parameter, which is chosen at random from the distribution:
\begin{equation}
P(\lambda)={1 \over {\sigma_{\lambda}(2 \pi)^{1/2}}} \exp\left[ {-\ln^{2}(\lambda/<\lambda>)}
\over{2 \sigma_{\lambda}^{2}} \right] {{d\lambda} \over {\lambda}},
\end{equation}
with $<\lambda>=0.05$ and $\sigma_{\lambda}=1.0$, 
\eg  Dalcanton \etal (1997) and references therein. The initial
dark halo profile is chosen such to have a central constant density core, as seen in present day
dark halos from dwarf to cluster scales, Firmani \etal  (2000). The formation of the disk
modifies the gravitational potential in the central regions, we include the reaction on the
dark halo to this process through an adiabatic invariance hypothesis 
(Flores \etal  1993; Hernandez \& Gilmore 1998).

Fig. 3 shows a typical PopIII proto-galaxy, having a total mass of 
$3\times 10^{8}M_{\odot}$, with a close-to-average
angular momentum, $\lambda=0.035$. The final rotation curve is given by the solid curve,
with the dotted line giving the surface density profile of the disk.

\begin{figure}
\label{fig3}
\psfig{file=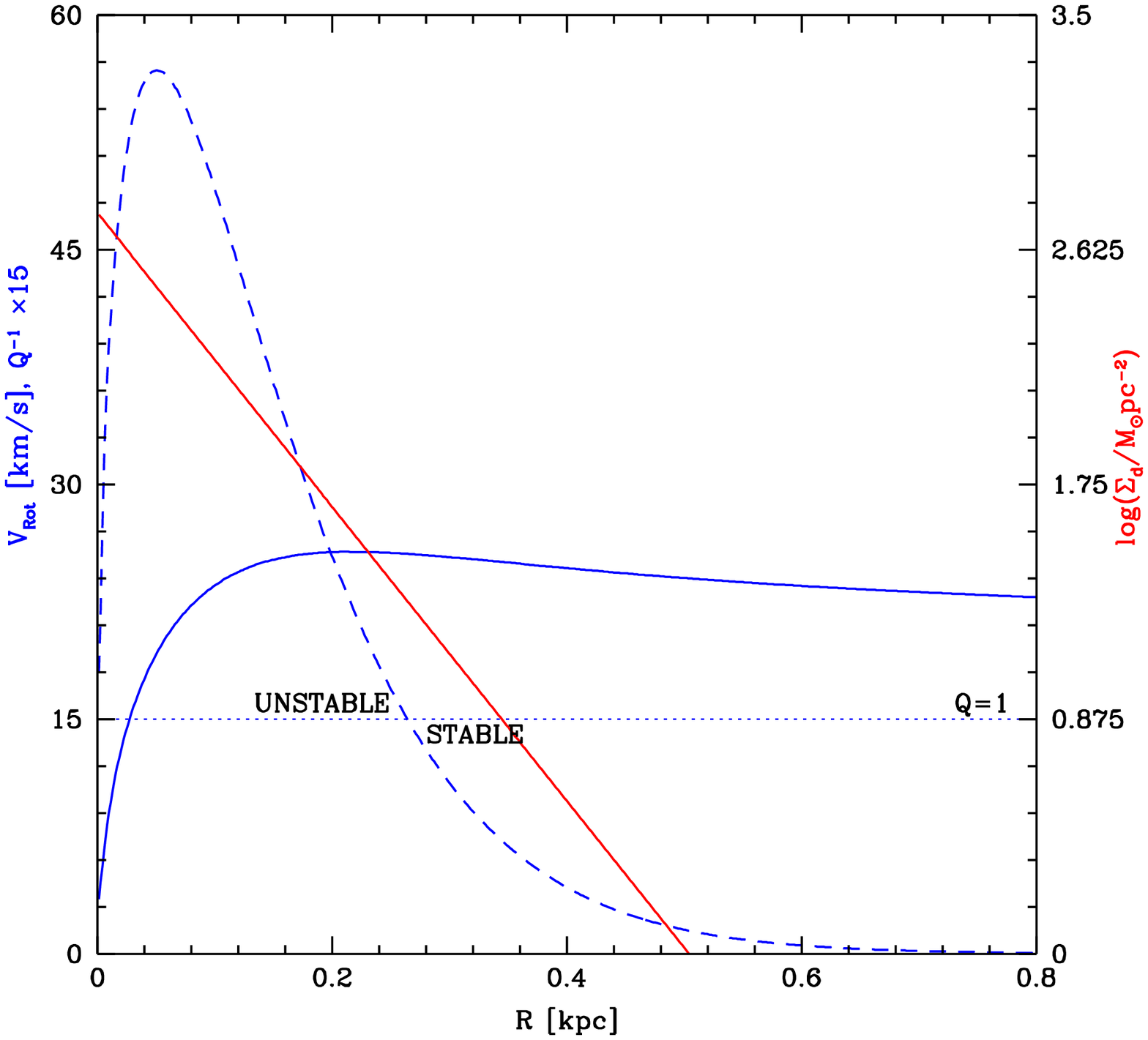,angle=0,width=9.0cm}
\caption{A typical PopIII galaxy, total mass 
$3\times 10^{8}M_{\odot}$, with a close to average
angular momentum, $\lambda=0.035$. The solid curve gives the rotation curve. The red line
represents the baryonic disk surface density profile; together with the estimated 
sound speed
in the disk, it gives the inverse of Toomre parameter, $Q^{-1}$, shown
by the dashed curve. The $Q=1$ instability threshold is shown by the dotted line,
with the intersection of this and the $Q^{-1}$ curve defining the stable and unstable
regions.
}
\end{figure}

\subsection{Disk Instabilities}

\begin{figure}
\label{fig4}
\psfig{file=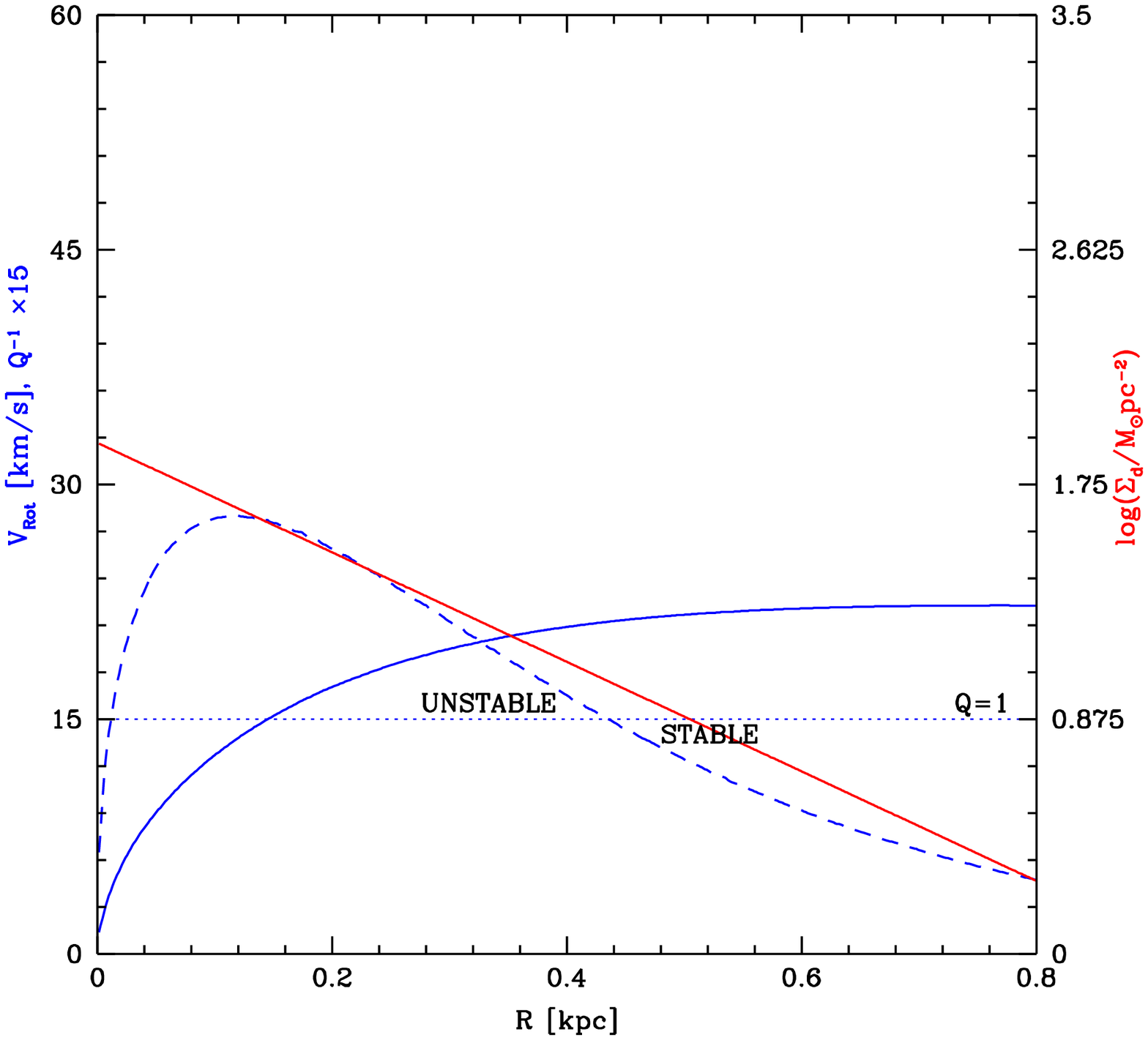,angle=0,width=9.0cm}
\caption{Same as Fig. 3, but for a higher value of the
 angular momentum, $\lambda=0.08$. The changes in
surface density profile, rotation curve (through the gravitational coupling of the disk 
and halo) and Toomre's parameter result in a larger instability radius, but a much reduced
fraction being susceptible to star formation. 
}
\end{figure}

The final temperature
of the gas in the disk is taken to be $300K$, as appropriate for gas where the main cooling
mechanism is H$_{2}$ (Ferrara 1998, Susa \& Umemura 2000, Machacek \etal 2000). This temperature determines $c_{s}$, the sound speed within the disk, 
which
together with $\kappa$, the epicycle frequency obtained from the detailed final rotation curve,
yields the Toomre's stability parameter for the disk:

\begin{equation}
Q={{c_{s}\kappa} \over {\pi G \Sigma}}.
\end{equation} 

In the regions where $Q>1$ the total tidal shears, together with the velocity dispersion, are
sufficient to stabilize the disk locally against its self gravity. These regions 
are therefore not subject
to star formation. The dashed curve in Fig. 3 gives the profile of $Q^{-1}$ in the disk,
showing the regions interior to the intersection with the $Q=1$ line to be unstable towards
star formation, in this case $85\%$ of the disk mass. Fig. 4 is analogous to Fig. 3, but for
 a system having a larger angular 
momentum,  $\lambda=0.08$. The changes in disk surface density and rotation curve imply
than now only a smaller fraction ($60\%$) of the disk will form stars. A value for the gas
temperature lower than the $300K$ we are assuming, due to the small presence of heavy 
elements in these objects, would
only increase slightly the value of the unstable fraction. Given that the unstable fractions 
for systems with typical values of $\lambda$ are already close to unity, this detail does not
affect our conclusions.

Finally, a gas to
star conversion efficiency
$f_\star =0.05$ is used to turn the total mass within the instability region into a final
mass of stars, for each object. Although we have calibrated this parameter from local
observations (Ciardi et al. 2000), since there is no firm theoretical prediction for
any evolution with redshift of it, we have not explored such possibility.

\subsection{Cloud Collisions and Collapse}

Although the above instability criterion might be satisfied, 
it does not necessarily follow that the considered PopIII objects will be able to form stars.
In fact, it is possible that cloud-cloud collisions might disrupt the gas clouds
formed by the gravitational instability before they have time to collapse (a point 
originally made by Kashlinsky \& Rees 1983). To explore this
possibility we calculate the time for collisions between clouds, $t_{coll}$ and for 
cloud collapse, $t_{ff}$. This last will be given by:
\be
t_{ff}\simeq \left( 1 \over G \rho \right)^{1/2},
\ee
where $\rho$ is the density of the gas clouds. The collisional time scale is 
\be
t_{coll}=(\pi {\cal N} \lambda_{J}^{2} c_{s})^{-1}, 
\ee
with ${\cal N}$ the number density of clouds, and $\lambda_{J}$ the Jeans  
length in the disk (in general a function of radius) which we take as a 
representative size for the clouds.
Taking ${\cal N}=f_c\rho / {M_{J}}$, where $f_c$ is the volume fraction of the disk 
filled with 
clouds, a filling factor, and the Jeans mass 
\be
M_J=\left( \pi \rho \over 6 \right) \left( \pi c_{s}^{2} \over {G \rho} \right)^{3/2},
\ee
since the Jeans length is:
\be
\lambda_{J}=\left( \pi c_{c}^{2} \over {G \rho} \right)^{1/2},
\ee
we can calculate the ratio of the collision time to the collapse time by substituting the values
above into eqs. 5 and 6 to yield,
\be
{t_{coll} \over t_{ff}}={2.9 \over f_c}.
\ee
Eq. 9 shows that even if the filling factor for the fragmenting disk were unity, which
is unlikely, the collapse time of the clouds is always shorter than the collision time,
at all radii, and independent of the value of $c_{s}$. This is a result of having taken
a constant value of $c_{s}$ throughout the disk, which is a consequence of H$_2$ being
the dominant cooling
mechanism operating in the primordial metallicity gas. As eq. 9 shows, for all the
scales of our problem, gravitational cloud collapse can proceed unimpeded by 
cloud-cloud collisions, which we shall not consider further.

\subsection{The First Stars Today}

Once the total gas mass turned into stars as a function of the metallicity 
has been determined, we use the IMF of
Larson (1998) to turn this into a total number of stars. This function offers a 
convenient parameterization of the IMF:
\begin{equation}
dN/dlog \hspace{2pt} m \propto (1+m/m_{s})^{-1.35}.
\end{equation}
In the above equation $m_{s}$ is a characteristic mass scale, of order 
$0.35M_{\odot}$ for a present day
solar neighbourhood IMF. This scale mass can be increased to explore the consequences
of a top heavy IMF applying to stars of metallicities much lower than solar, as the ones we
are treating. 

We combine this sampling of the IMF using $m_{s}=0.35 M_\odot$ with the 
most recent Padova stellar models of primordial 
metallicities of Girardi \etal  (2000) to exclude those stars which due to
their age have by now evolved off the tip of the RGB, and are hence no longer visible. In this
way we obtain a sample of present day stars.
Sampling the IMF from a fixed lower luminosity up to the tip of the RGB of the 
theoretical isochrones allows us to obtain the number of stars visible today upwards of
a certain limiting luminosity, as a function of metallicity. Our final samplings include
only stars between $0.6 M_{\odot}$ and the tip of the RGB of the Padova isochrones 
(of around $1.0M_{\odot}$), which is the range  over which stellar models were calculated.

As the total mass in PopIII objects is high,  
both the PopIII halos mass functions and the $\lambda$ distributions
are sampled densely, yielding a low spread for the predicted final number of stars.

To summarize, each of the PopIII object selected from the mass function of eq. 1 
is carefully modeled as a mini galaxy, with a value of $\lambda$ selected at random
from eq. 3. The resulting system is then treated as a mini galaxy where the cooling
of the baryons leads to disk formation; subsequently a disk instability criterion is used to obtain the
total mass in stars contributed by each object. Finally, through the use of an IMF and
new low metallicity stellar models, this total mass is turned into a number of stars 
visible today. This is repeated until the total PopIII
object mass equals that within the bounded region described in Section 2, for a given
limiting redshift, which corresponds to a fixed maximum metallicity. This process is repeated
for different limiting redshifts, to construct a total mass distribution function for the 
low metallicity stars for the present day MW. The curve in Fig. 5 shows the number of MW stars 
with mass in the above range and having metallicities lower than 
a given value.

\section{COMPARISON WITH DATA}

To compare our results with observations we take the sample of Ryan \& Norris (1991)
of low metallicity halo stars in the local disk. They include a magnitude limited
sample, (their NLTT sample) which is complete down to $0.6 M_{\odot}$ and $V=13.0$, 
the same limit we use to sample our IMF.
From the theoretical isochrones, this corresponds to a distance of 290 pc, which we take
as the radius of the sample. In that sample 5 stars were found with $[Z/Z_{\odot}]<-3.0$.
To obtain the total number of stars in the Galactic halo at this metallicity, we
must introduce an assumption regarding the distribution of these stars.

\begin{figure}
\label{fig5}
\psfig{file=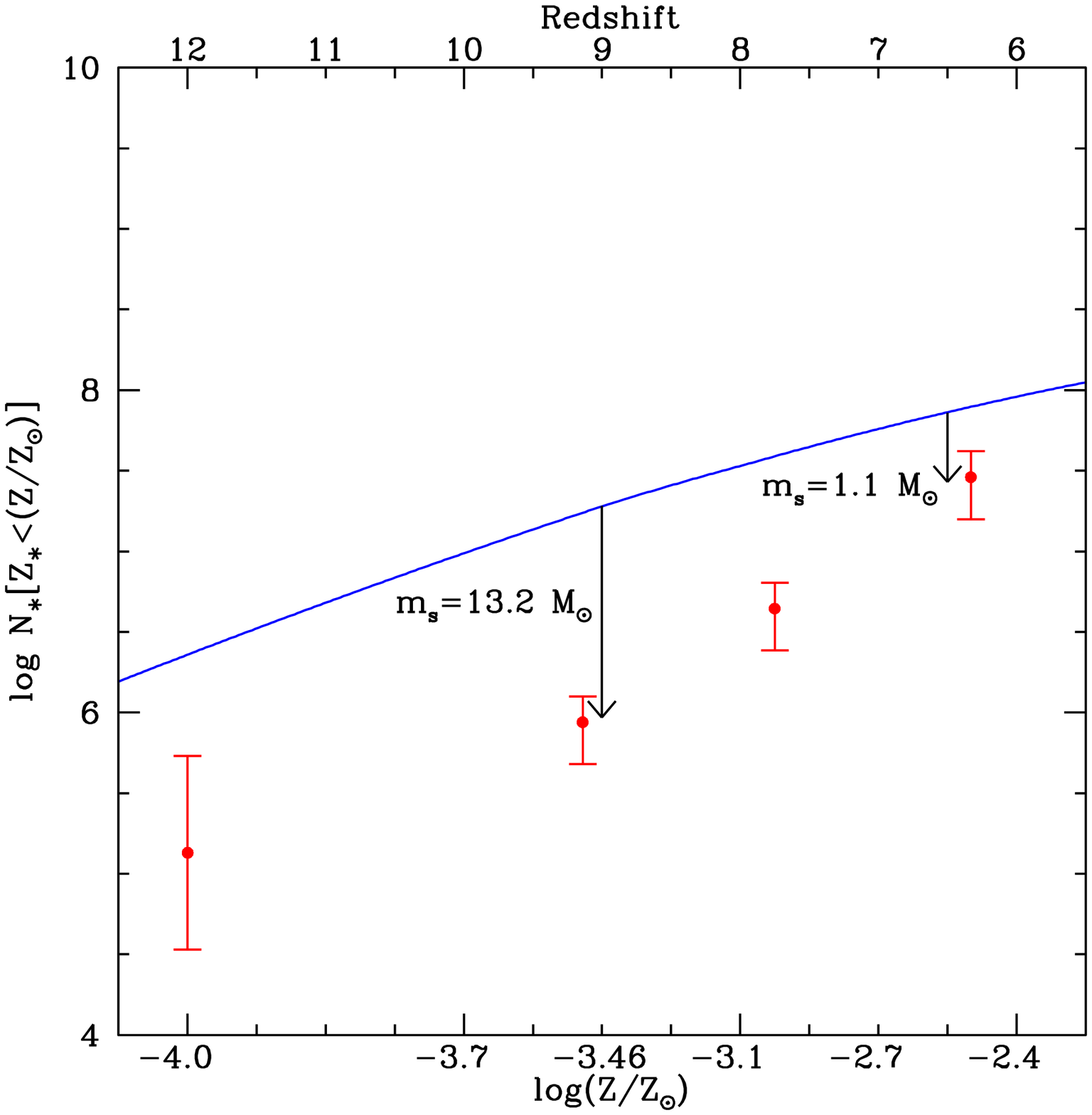,angle=0,width=9.0cm}
\caption{Average number of PopIII stars with metallicity lower than
$[Z/Z_\odot]$ expected in MW sized galaxies today, for a solar
neighbourhood IMF (solid curve). The dots with error bars show the 
observations of Ryan \& Norris (1991) and Beers \etal  (1998), assuming 
an isothermal distribution for the stars in the halo (see text).
The labeled arrows show the values of $m_{s}$ required to reconcile model and
data, at $z=6.5$ and $z=9.0$.
}
\end{figure}

If the PopIII stellar systems are disrupted while they are incorporated into the growing
MW, their stars will today be found almost exclusively in the Galactic halo, while
if the initial stellar systems form tight clusters, they might survive disruption and
through dynamical friction end up in the galactic bulge. Which of these two processes
dominates will depend on the mass ratio of the mergers through which the PopIII systems
reach the MW. If these mass ratios are close to one, disruption will be important,
while if they are accreted late into a fully formed MW PopIII objects might
remain bound. From N-body simulations White \& Springel (2000) have argued that most of the lowest
metallicity stars in a present day galaxy will be found in the stellar halo, and not in the bulge.
We therefore make the assumption that the lowest metallicity stars share the isothermal
distribution of the dark halo, out to a distance of $100$ kpc, representative of the total
extent of the halo. The actual density profile of these objects might more closely resemble that
of the old globular cluster population, having an $R^{-3}$ profile. Fortunately, fixing the 
normalization point at the solar radius (8.5 kpc) yields total numbers which are largely 
independent of this change in the assumed profile, e.g. for core radii in the $R^{-3}$
profile going from 0.2 to 2 kpc (a core radius must be introduced in the $R^{-3}$ profile to avoid
the central divergence), the total numbers of stars change by a factor 1.24 to 0.34, respectively,
small compared to our error bars in figures 5 and 6.
It is worth pointing out that the assumption of a particular density profile for the current
low metallicity halo stars is not 
intrinsic to our results, but becomes necessary only for data comparison purposes.

\begin{figure}
\label{fig6}
\psfig{file=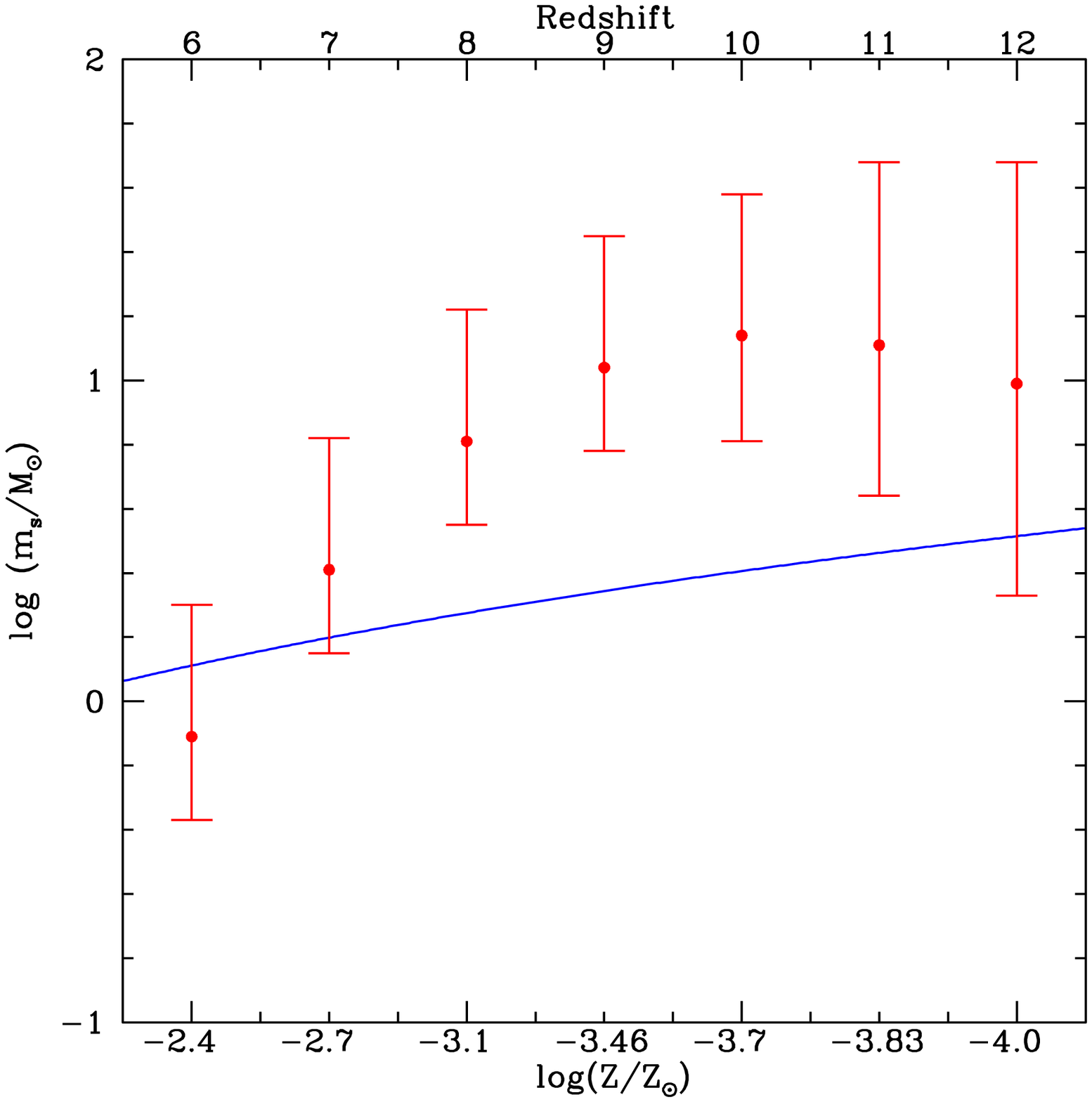,angle=0,width=9.0cm}
\caption{Inferred values of $m_{s}$ as a function of redshift (points with error bars).
The solid curve gives the redshift evolution of the Jeans mass of cold star forming clouds, 
identified with $m_{s}$, resulting from the temperature evolution of the cosmic microwave 
background (CMB).
}
\end{figure}

Taking the Ryan \& Norris (1991) data plus the hypothesis of an isothermal 
distribution discussed above, we derive the total number of stars with 
$[Z/Z_{\odot}]<-3.0$ in the Galaxy as 
$4.43 \times 10^{6}$. To obtain numbers at other metallicities we use the relative
metallicity distribution function of the more recent Beers \etal  (1998) work, which 
combines several new samples. These observations are shown in Fig. 5 by the points
with error bars, which reflect the uncertainties in the normalization. 

The stellar halo might not extend as far out as $100$~kpc, but some PopIII stars might be 
found in the bulge, so the interpretation of data used in Fig. 5 is probably a lower limit. 
On the other hand, given the uncertainties in the primordial IMF,
our prediction is a strict upper limit. 

It is clear that our results fall somewhat above the observational measurements, implying
that our assumption of a constant $m_{s}=0.35 M_\odot$, essentially a present day solar neighbourhood
IMF, is not valid. Reconciling our estimates with the observations requires not only the 
assumption of a higher $m_{s}$ in the past, but throughout the redshift range studied,
an increasing trend for this value. Note that our derived numbers of PopIII stars scale
linearly with $f_{\star}$

\begin{figure}
\label{fig7}
\psfig{file=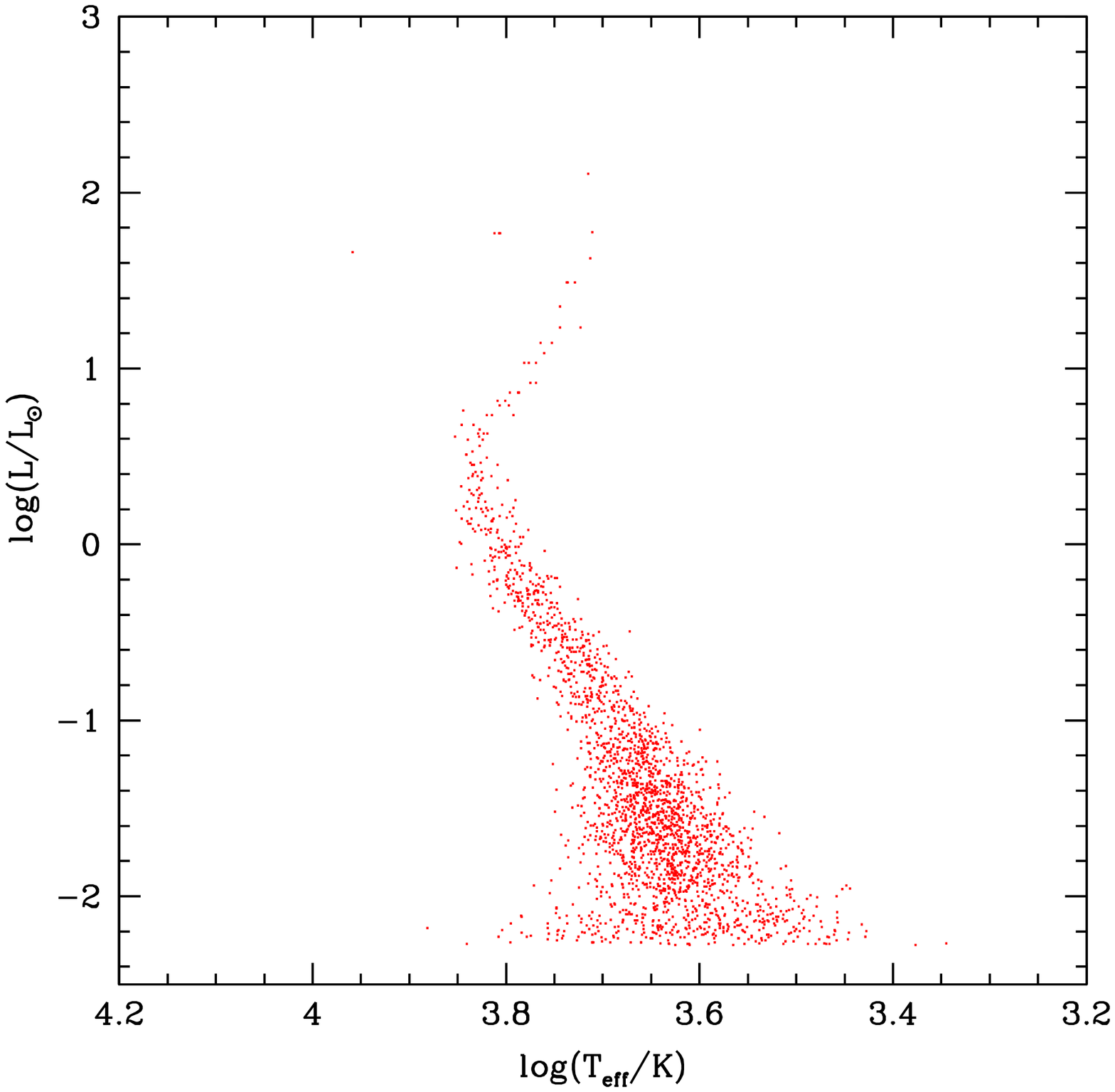,angle=0,width=9.0cm}
\caption{HR diagram for 2286 extremely low metallicity stars.
 Accurate isochrones for these stars 
are used to obtain good estimates of their present observational characteristics, which can
be used to provide guidelines of their current photometrics and hence aid detection strategies.
}
\end{figure}

Several authors, \eg  Larson (1998) have pointed out that the increase of the CMB temperature with
redshift alone, implies an increase in the Jeans mass for the star forming clouds, once one
is beyond $z=2$, where the CMB temperature equals the $\approx 8$~K of cold star forming
clouds. It has therefore been suggested that at high redshift the IMF should be 
increasingly weighted towards larger stellar masses. The labeled arrows in Fig 5. show the
values of $m_{s}$ which we require at $z=6.5$ and $z=9.0$, to match the observed points.

In Fig. 6 we show the values of $m_{s}$ which we infer for the IMF of PopIII objects, 
by imposing that model and observations agree.
The solid curve indicates the redshift evolution of the Jeans mass in star
forming clouds, which we identify with $m_{s}$. Although for $z<9$ our results are consistent
with the increase in $m_{s}$ coming solely from that in the Jeans mass due to the CMB, 
the trend for $6<z<9$ is clearly for a rise in $m_{s}$ beyond what this effect can account for.
In fact, as Larson (1998) points out, this is only the {\it minimum} increase to be expected,
as other heating sources, \eg  massive stars themselves or a UV background, 
were likely to be present. Furthermore, a reduced opacity consistent with these low metal abundances
would have the effect of skewing  the IMF towards larger masses, thus 
operating in the same direction.

For $z>9$, the values of the characteristic mass appear to stabilize around the value
$m_s = 10-15 M_\odot$, indicative of a very top heavy IMF at these early epochs,  
although the increase of our error bars makes determining trends in this redshift
range harder.

Finally,  Fig. 7 presents an HR diagram containing over 2000 stars, produced from the 
low metallicity
stellar models used, convolved with errors appropriate to HST observations of halo stars 
(Gallart \etal  1999; Hernandez \etal 2000), for 
$m_{s}=5.0 M_{\odot}$. The inclusion of detailed atmosphere calculations and spatial information
to yield detailed observational predictions will form part of a forthcoming paper.

\section{SUMMARY}         

From our study we conclude that:

\begin{itemize}
\item If the standard hierarchical merger scenario is correct, in the MW        
one should not expect
to find stars having metallicities lower than
$[Z/Z_{\odot}]=-4.3$, consistent with current observational limits. 

\item We infer the IMF of PopIII stars to have been strongly weighted towards high masses,
and increasingly more so in the range $6<z<9$, this levelling off at 
$z\simgt 9$ at a characteristic mass scale $m_{s}=10-15 M_{\odot}$.

\item Observational limits on the numbers and metallicity distribution of very low metallicity 
stars in the MW can
place sharp restrictions on the IMF of the first stars. This can be achieved through 
detailed modelling of 
the full range of phenomena giving rise to such stars, of the type attempted here,
thus necessarily establishing a connection with the underlying structure formation scenario.

\end{itemize}

\section{ACKNOWLEDGMENTS}
The authors wish to thank the referee, Volker Bromm, for his careful reading of the manuscript
and helpful comments which improved the presentation of this paper, as well as Richard Larson
for a detailed look at a first draft and Leo Girardi for making his new isochrones available 
electronically. This work was completed as one of us (AF) was a Visiting Professor at
the Center for Computational Physics, Tsukuba University, whose support
is gratefully acknowledged.

\end{document}